\documentclass[runningheads]{llncs}
\usepackage[T1]{fontenc}
\usepackage{graphicx}
\usepackage{todonotes}
\usepackage{listings}
\usepackage{hyperref}
\usepackage{xcolor}

\begin{document}

\title{CHAD-KG: A Knowledge Graph for Representing Cultural Heritage Objects and Digitisation Paradata}

\titlerunning{CHAD-KG: A Graph for Heritage Objects and Paradata}

\author{
Sebastian Barzaghi\inst{1}\orcidID{0000-0002-0799-1527} 
\and
Arianna Moretti\inst{2}\orcidID{0000-0001-5486-7070}
\and
Ivan Heibi\inst{2}\orcidID{0000-0001-5366-5194} 
\and
Silvio Peroni\inst{2}\orcidID{0000-0003-0530-4305}
}

\authorrunning{S. Barzaghi et al.}

\institute{
Department of Cultural Heritage, University of Bologna, Via Degli Ariani, 1, 48121, Ravenna, RA, Italy 
\and
Department of Classical Philology and Italian Studies, University of Bologna, Via Zamboni, 32, 40126, Bologna, BO, Italy
}

\maketitle              

\begin{abstract}
This paper presents CHAD-KG, a knowledge graph designed to describe bibliographic metadata and digitisation paradata of cultural heritage objects in exhibitions, museums, and collections. It also documents the related data model and materialisation engine. Originally based on two tabular datasets, the data was converted into RDF according to CHAD-AP, an OWL application profile built on standards like CIDOC-CRM, LRMoo, CRMdig, and Getty AAT. A reproducible pipeline, developed with a Morph-KGC extension, was used to generate the graph. CHAD-KG now serves as the main metadata source for the Digital Twin of the temporary exhibition titled \emph{The Other Renaissance - Ulisse Aldrovandi and The Wonders Of The World}, and other collections related to the digitisation work under development in a nationwide funded project, i.e. Project CHANGES (\url{https://fondazionechanges.org}). To ensure accessibility and reuse, it offers a SPARQL endpoint, a user interface, open documentation, and is published on Zenodo under a CC0 license. The project improves the semantic interoperability of cultural heritage data, with future work aiming to extend the data model and materialisation pipeline to better capture the complexities of acquisition and digitisation, further enrich the dataset and broaden its relevance to similar initiatives.


\end{abstract}

\section{Introduction}\label{intro}

In recent years, there has been a growing interest in developing and applying Semantic Web technologies to create Knowledge Graphs (KGs), that leverage knowledge representation to improve the application of FAIR principles~\cite{wilkinson_fair_2016} and reproducibility for effective scientific data management and use~\cite{stocker_skg4eosc_2022}. Interest in applying semantic technologies has also grown within the humanities and cultural heritage fields. This trend is driven, on the one hand, by the nature of their data—heterogeneous, multi-topical, multilingual, large-scale, highly contextualised by time and place, increasingly collaborative in creation, and often requiring user-specific personalisation~\cite{gualandi_what_2022}~\cite{hyvonen_knowledge-based_2019}. On the other hand, it reflects a broader, ongoing shift within the humanities towards embracing the FAIR and Open Science paradigms~\cite{grant_reusable_2023}~\cite{harrower_sustainable_2020}~\cite{barbuti_creating_2020}~\cite{larsson_developing_2025}~\cite{toth-czifra_risk_2019}~\cite{poljak_bilic_fairness_2024}. Indeed, by facilitating the representation and sharing of data on the web in the form of machine-actionable Linked Open Data (LOD), KGs and services leveraging them hold the potential to positively impact how humanistic and cultural data are generated, curated, accessed, and reused~\cite{mero241_clariah_2020}.

While recent work has demonstrated the potential of KGs in representing cultural heritage data, several challenges remain underexplored, particularly in relation to the processes that support the digitisation of Cultural Heritage Objects (CHOs)~\cite{ioannides_integrating_2025}. Most existing KGs in the cultural heritage domain focus on descriptive assertions about artefacts, often overlooking the dynamic and procedural metadata (paradata) that document how these artefacts are acquired, digitised, and curated~\cite{desul_semantic_2023}. Moreover, despite noteworthy developments on that front, there is still too little attention given to standardised approaches to ensure that such data is FAIR-compliant. A further challenge concerns the degree of human involvement in managing KGs: although a wide range of different approaches emerged recently, from automatic knowledge extraction based on Deep Learning algorithms and Large Language Models~\cite{khorashadizadeh_research_2024}~\cite{zhong_comprehensive_2023}, to “human-in-the-loop” methodologies emphasising more participation of the human element~\cite{rahman_knowledge_2024}~\cite{ristoski_large-scale_2020}, there is still need for close collaboration between technical experts and domain specialists, which needs to be continuous to be fruitful~\cite{angelis_chekg_2024}~\cite{jain_domain-specific_2020}. These limitations become particularly evident in projects aiming to combine 3D digitisation, metadata enrichment, and virtual access to collections, where sustainability, interoperability, transparency and communicability of the underlying workflows are critical~\cite{barzaghi_proposal_2024}.

One key to addressing these challenges lies in effective semantic data modelling, which serves as the foundation for any KG~\cite{hao_construction_2021}, as it provides the formal structure that underpins the interpretation, integration, and reuse of data~\cite{perez-arriaga_construction_2018}. In the context of cultural heritage, this means not only capturing descriptive metadata of CHOs (such as their title, identifiers, relevant dates, etc.), but also contextual information such as provenance, digitisation events, and the relationships between them and their digital facsimiles~\cite{doerr_ontologies_2009}. Leveraging existing models, such as ontologies and controlled vocabularies, ensures semantic alignment across domains and institutions, which is particularly important for ensuring data interoperability~\cite{guizzardi_ontology_2020} and reusability~\cite{thanos_research_2017}, even in multidisciplinary and collaborative settings. Thus, robust and extensible semantic data models are essential for ensuring long-term interpretability and relevance, as a KG can grow and be integrated with other datasets aligned to the same vocabularies.

Equally important is materialisation, the process of constructing KGs from various data sources by converting data into Resource Description Framework (RDF) triples~\cite{van_assche_balancing_2022}. This involves not only technical conversion but also decisions about how to semantically align disparate data formats and content with the target model. In particular, this step enables the integration of heterogeneous cultural heritage data into a unified, semantically rich graph that can be queried, analysed, and reused.

In this paper, we describe the KG produced so far as one of the outcomes of the CHANGES Project (\url{https://fondazionechanges.org}), a nationwide project funded by the European Union and dedicated, among other activities, to the use of digital technologies to support conservation, valorisation and public engagement with cultural heritage. First, we present the \emph{Cultural Heritage Acquisition and Digitisation Knowledge Graph} (CHAD-KG), a KG containing descriptive metadata and digitisation paradata of a series of cultural objects that were part of a temporary exhibition, closed in May 2023, entitled \emph{The Other Renaissance: Ulisse Aldrovandi and the wonders of the world} (\url{https://site.unibo.it/aldrovandi500/en/mostra-l-altro-rinascimento}). Second, the application profile CHAD-AP -- introduced in~\cite{barzaghi_developing_2025} and developed following the SAMOD methodology~\cite{peroni_simplified_2017} -- is described. Third, a materialisation pipeline, based on Morph-KGC~\cite{arenas-guerrero_morph-kgc_2024}, is introduced and contextualised through its application for producing RDF data from the objects’ metadata and digitisation paradata contained in two tabular datasets, resulting in a KG that adheres to the developed application profile.

The rest of the paper is structured as follows. Section~\ref{related} illustrates existing works in constructing datasets, ontologies, and conversion mechanisms to better contextualise our work. Section~\ref{resource} describes the KG produced as a result of its conceptualisation (Section~\ref{model}), materialisation (Section~\ref{materialisation}) and publishing (Section~\ref{publishing}). Section~\ref{discussion} discusses and evaluates how the KG was developed and tested, alongside the model and the pipeline. Finally, Section~\ref{conclusion} concludes the paper, highlighting strengths and limitations of our approach and illustrating possible future directions.

\section{Related work}\label{related}

%For the literature review collection and analysis we decided to focus on projects as close as possible to our scope, thus opting to target semantic metadata production pipelines of cultural heritage digitisation processes and studying the interconnection between the technical and procedural choices taken as a whole. Since the set of requirements was broad and specific, the selection was extended to projects on which it was possible to at least partially analyse the organicity of the choices concerning the semantic data materialisation workflow, emphasising the points of internal coherence aimed at the production of a KG.

Many KGs have been produced over the past years to support different operations on data about cultural heritage and its digitisation, ranging from data representation, extraction and analysis to knowledge enrichment, retrieval, and publication on the web~\cite{pellegrino_move_2022}. Such datasets often consist of RDF triples modelled according to well-known ontologies already in use within the domain for structuring CHOs’ descriptions and information about their digitisation lifecycle, such as the Provenance Ontology (PROV-O) (\url{http://www.w3.org/ns/prov})~\cite{lebo_prov-o_2013}, the CIDOC Conceptual Reference Model (CIDOC-CRM) (\url{http://www.cidoc-crm.org/cidoc-crm/})~\cite{doerr_cidoc_2003} and the Europeana Data Model (EDM) (\url{http://www.europeana.eu/schemas/edm/})~\cite{haslhofer_dataeuropeanaeu_2011}. Examples of use of these ontologies include the work by~\cite{homburg_metadata_2021}, discussing the development of an ontology based on PROV-O for describing 3D CHOs capturing and processing workflow, and~\cite{hiebel_ontological_2021}, using CIDOC-CRM and its extensions to generate a unified representation of entities related to the excavations at Tell el-Daba, Egypt, including physical and digital resources, sites, and process documentation.

With respect to KGs produced within projects focused on cultural heritage digitisation, many cases focus on the semantic modelling of objects and activities based on CIDOC-CRM. For example,~\cite{padfield_semantic_2019} describe a case study on semantically modelling cultural heritage data from the National Gallery in London. Their approach uses a custom extension of CIDOC-CRM and Getty vocabularies, along with a data processing pipeline that aggregates and provides access to various data managed by the Gallery through a dedicated API. The resulting JSON data is then mapped to the developed ontology and converted into an RDF KG using a set of PHP functions. In~\cite{schleider_silknow_2021}, the authors describe how they leveraged Semantic Web technologies to generate a KG to model data about European silk heritage from the 15th to the 19th century within the SILKNOW project. The KG data model is based on CIDOC-CRM, the CRM Scientific Observation model (CRMsci) (\url{https://cidoc-crm.org/extensions/crmsci/})~\cite{doerr2014crmsci} and the CRM Digital extension (CRMdig) (\url{http://www.cidoc-crm.org/extensions/crmdig/})~\cite{doerr2011crmdig}, with the addition of a custom CRM extension for dealing with information related to textile artefacts and a custom multilingual SKOS thesaurus to systematise silk heritage terminology. With respect to the data itself, the SILKNOW knowledge graph consists of publicly available datasets from museum collections that are part of the project. The data processing pipeline includes a NodeJS-based crawler that collects metadata records from various museums’ APIs or websites, and a converter written in Java to parse, clean and harmonise metadata records into an unified KG based on the SILKNOW data model. Similarly, ~\cite{yang_knowledge_2023} describe a method for representing data through a KG to support semantic 3D modelling of Chinese grottoes. The data model underlying the knowledge representation incorporates CIDOC-CRM and GeoSPARQL~\cite{battle_enabling_2012} to describe both geospatial and cultural heritage data. To construct the KG, they use tools typically used for Natural Language Processing tasks, such as OpenNRE~\cite{han_opennre_2019} and CRF++, to extract knowledge from monographs and academic papers, formalise it according to the data model, and store it in a Neo4j graph database. In their work, ~\cite{carriero_arco_2019} present ArCo, a KG dedicated to describing Italy’s cultural heritage data by leveraging catalog records extracted from the General Catalog of the Italian Ministry of Cultural Heritage (MiBAC) and converted into RDF triples that follow a series of ontologies based on Ontology Design Patterns (ODP)~\cite{gangemi_ontology_2009} and aligned with a number of existing models, including CIDOC-CRM, EDM, DBpedia~\cite{lehmann_dbpedia_2015}, Wikidata~\cite{vrandecic_wikidata_2014} and Geonames. The RDF data are generated via a RDFizer tool, written in Java, that relies on XSLT to take XML catalogue records as input, process them, and transform them into a unified RDF dataset. Finally, ~\cite{peroni_food_2016} illustrate how they transformed data from Italian policy documents regulating agricultural products into a LOD dataset. Within the FOod in Open Data (FOOD) project, they developed a modular ontology using established design patterns and aligned it with vocabularies like AGROVOC~\cite{rajbhandari_agrovoc_2012} and DBpedia. Over 800 documents were processed, with partial automation for DOCX files and manual extraction from PDF documents. The resulting RDF datasets include structured data on product types, raw materials, production areas, and quality characteristics, all released under an open license.

\section{The Cultural Heritage Acquisition and Digitisation Knowledge Graph (CHAD-KG)}\label{resource}

The main contribution of this paper is the Cultural Heritage Acquisition and Digitisation Knowledge Graph (CHAD-KG)~\cite{moretti_2025_15102846}. This KG contains the descriptive metadata and digitisation paradata about the CHOs, displayed in a temporary museum exhibition dedicated to Ulisse Aldrovandi, an Italian naturalist lived between 1522 and 1605, recognised as one of the fathers of natural history studies. The exhibition -- ended definitively on May 28th, 2023 -- hosted a vast range of different items, such as books, maps, specimens, and scientific instruments, each documented by the museum with its own set of descriptive metadata. In an endeavour spanning the last couple of years, the exhibition objects were digitised into a collective Digital Twin, to recreate a version of the exhibition that is both scientifically documented and reproducible, and usable by different users~\cite{balzani_saving_2024}. 

CHAD-KG is the outcome of a digital humanities project built on an integrated process. Its focus extended beyond the production of an RDF graph per se, but the exposure of readily usable data, designed to be maintained and updated over time, as the result of a FAIR and replicable methodology. Thus, this research approach treats the KG as part of an ongoing, collaborative process involving experts from various fields, which calls for modelling and tools that are suitable for a range of users, not just those with Semantic Web expertise.  

Consistently, the requirements we elected to fulfill in producing the semantic graph include: (1) interoperability with existing standards for LOD; (2) possibility of future extension of both the graph and the model it is based on to handle needs not manifested in the first case study; (3) possibility of communicating modelling choices to all team members; (4) achievement of a sufficient degree of generality to guarantee reproducibility of the methodology and easy integration with data produced in subsequent case studies; (5) respect for the principles of FAIRness and compliance with the Open Science ethos.

CHAD-KG currently serves as the source of truth for the information used to describe the CHOs and their digitised replicas, or Digital Cultural Heritage Objects (DCHOs). In its current version, openly available on Zenodo~\cite{moretti_2025_15102846} under a CC0 1.0 license, CHAD-KG comprises a total of 52,080 triples describing 14,506 entities, including the CHOs in the exhibition -- expressed according to the IFLA Library Reference Model (IFLA LRM)~\cite{zumer_ifla_2018} -- and all the accompanying entities that are part of the CHOs’ history, context and digitisation process (people, organisations, activities, tools, licenses, and so on).

In the rest of this section, we introduce our approach for conceptualising (Section~\ref{model}), materialising (Section~\ref{materialisation}) and publishing (Section~\ref{publishing}) CHAD-KG.

\subsection{Data Model}\label{model}

As a first step towards the creation of a KG, we conceptualised the domain and developed an appropriate semantic data model. More specifically, the data contained in the KG is structured around the Cultural Heritage Acquisition and Digitisation Application Profile (CHAD-AP) (\url{https://w3id.org/dharc/ontology/chad-ap})~\cite{barzaghi_developing_2025}, an OWL ontology for describing CHOs and the processes of acquisition and digitisation in the cultural heritage domain as structured, machine-actionable data. To develop CHAD-AP, the Simplified Agile Methodology for Ontology Development (SAMOD)~\cite{peroni_simplified_2017} was followed, in a total of 11 iterations, each documented on its GitHub repository (\url{https://github.com/dharc-org/chad-ap/}). Overall, the current version (2.0.1) of CHAD-AP reuses a total of 22 classes, 56 properties, and 86 individuals.

CHAD-AP is composed of two main modules. One is the Object Module (OM), based on CIDOC-CRM (version 7.1.3) and its extension Object-Oriented Library Reference Model (LRMoo, version 1.0) (\url{http://iflastandards.info/ns/lrm/lrmoo/})~\cite{riva_lrmoo_2022}, and used to describe characteristics and contextual information of CHOs. The other is the Process Module (PM), mainly based on CRMdig (version 4.0), and used to describe the process of acquiring and digitising CHOs to produce new DCHOs. Both modules also reuse Getty’s Art and Architecture Thesaurus (AAT) (\url{http://vocab.getty.edu/aat/})~\cite{harpring_development_2010} to describe specific aspects of data expressed through well-defined, standardised terminology, such as the CHO’s type, actors’ roles in its creation, the activities involved in the digitisation process, and so on.

\begin{figure}[h!]
\includegraphics[width=\linewidth]{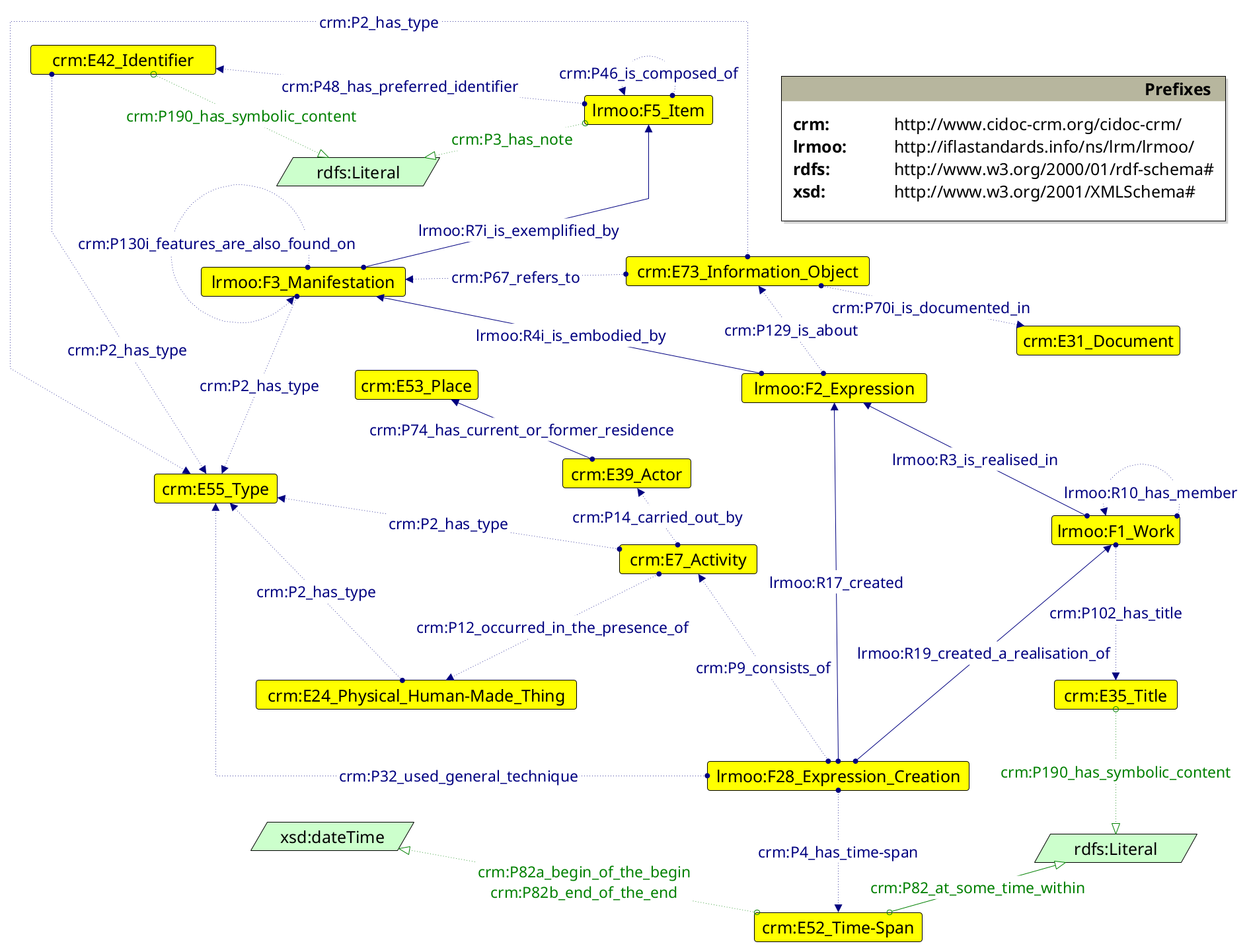}
\caption{A diagram of the CHAD-AP Object Module (OM).} \label{object-model}
\end{figure}

On the one hand, OM (Figure~\ref{object-model}) allows representing CHOs as constructs with their metadata distributed along four conceptual layers. The first layer is the Work (the essence of the object, modelled as \texttt{lrmoo:F1\_Work}), which is characterised by one or more titles (each an instance of \texttt{crm:E35\_Title}), as well as relationships with other Works. The second layer is the Expression (the intellectual realisation of the Work, modelled as \texttt{lrmoo:F2\_Expression}), tied to the set of entities that contributed to the CHO’s creation (usually instances of \texttt{crm:E7\_Activity}) as well as its content (including its subjects as instances of \texttt{crm:E73\_Information\_Object}). The third layer is the Manifestation (the embodiment of an Expression in a format, modelled as \texttt{lrmoo:F3\_Manifestation}), which describes the CHO’s type (\texttt{crm:E55\_Type}) and its related license (a semantic pattern centered around \texttt{crm:E73\_Information\_Object}). The fourth layer is the Item (the actual, physical CHO exemplar, modelled as \texttt{lrmoo:F5\_Item}), accompanied by descriptive labels and eventually linked to information related to its curation and conservation.

\begin{figure}[h!]
\includegraphics[width=\linewidth]{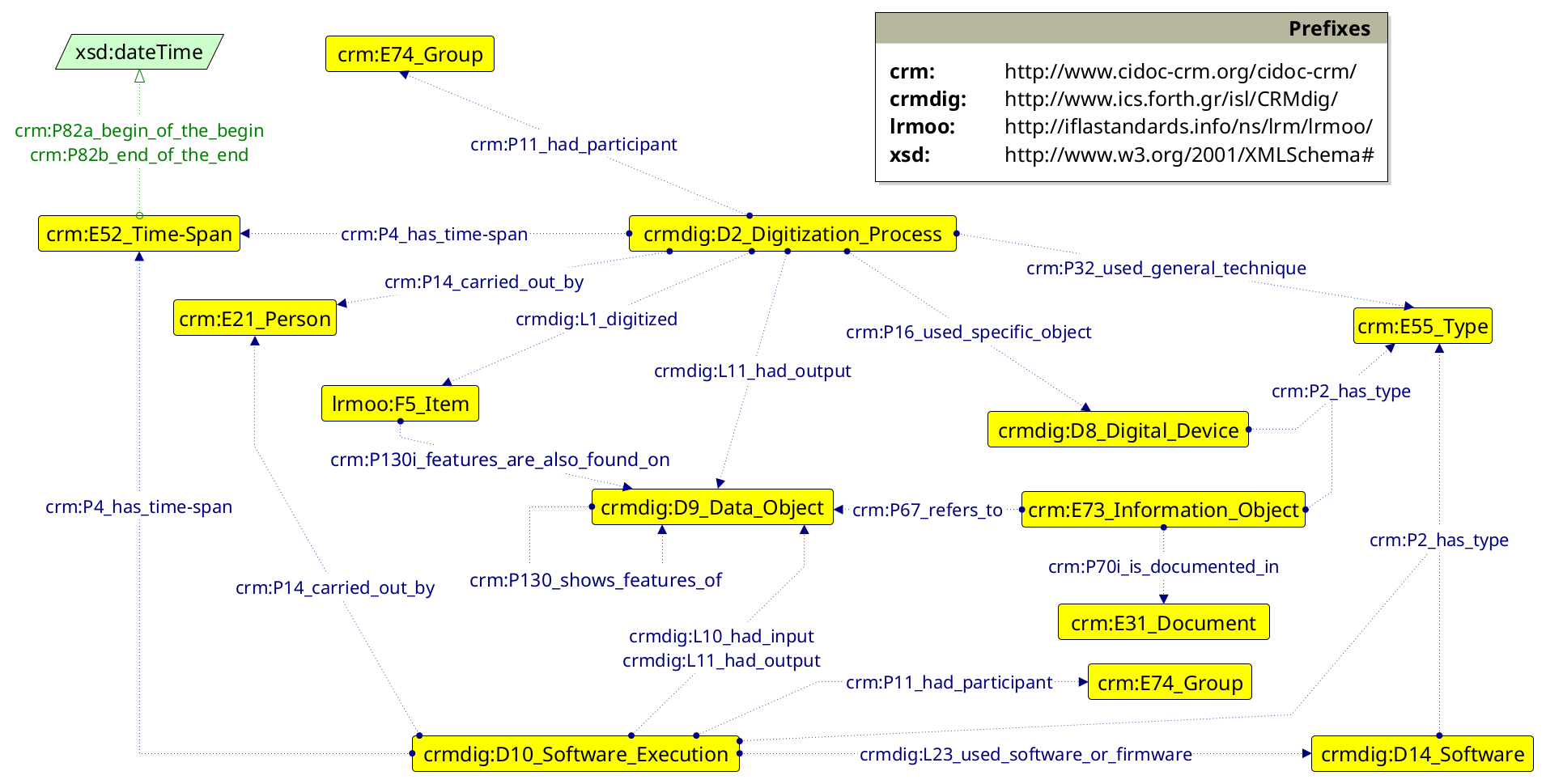}
\caption{A diagram of the CHAD-AP Process Module (PM).} \label{process-model}
\end{figure}

On the other hand, PM (Figure~\ref{process-model}) focuses on representing a digitisation workflow as a sequence of activities. The first (\texttt{crmdig:D2\_Digitization\_Process}) is the acquisition and involves the CHO at the Item level and produces the first, raw digitisation data (\texttt{crmdig:D9\_Data\_Object}). The others, collectively referred to as software activities (\texttt{crmdig:D10\_Software\_Execution}), represent actions like processing, modelling, optimisation, and so on. Both acquisition and software activities involve other entities, including: the 3D data (\texttt{crmdig:D9\_Data\_Object}) passing from one activity to the other as either input or output; the people (\texttt{crm:E21\_Person}) and organisations (\texttt{crm:E74\_Group}) that carry out the activities; the techniques (\texttt{crm:E55\_Type}) and tools (\texttt{crmdig:D8\_Digital\_Device} and \texttt{crmdig:D14\_Software}) involved; and the activities’ temporal information (\texttt{crm:E52\_Time-Span}).

\subsection{Materialisation Pipeline}\label{materialisation}

\begin{figure}[h!]
\includegraphics[width=\linewidth]{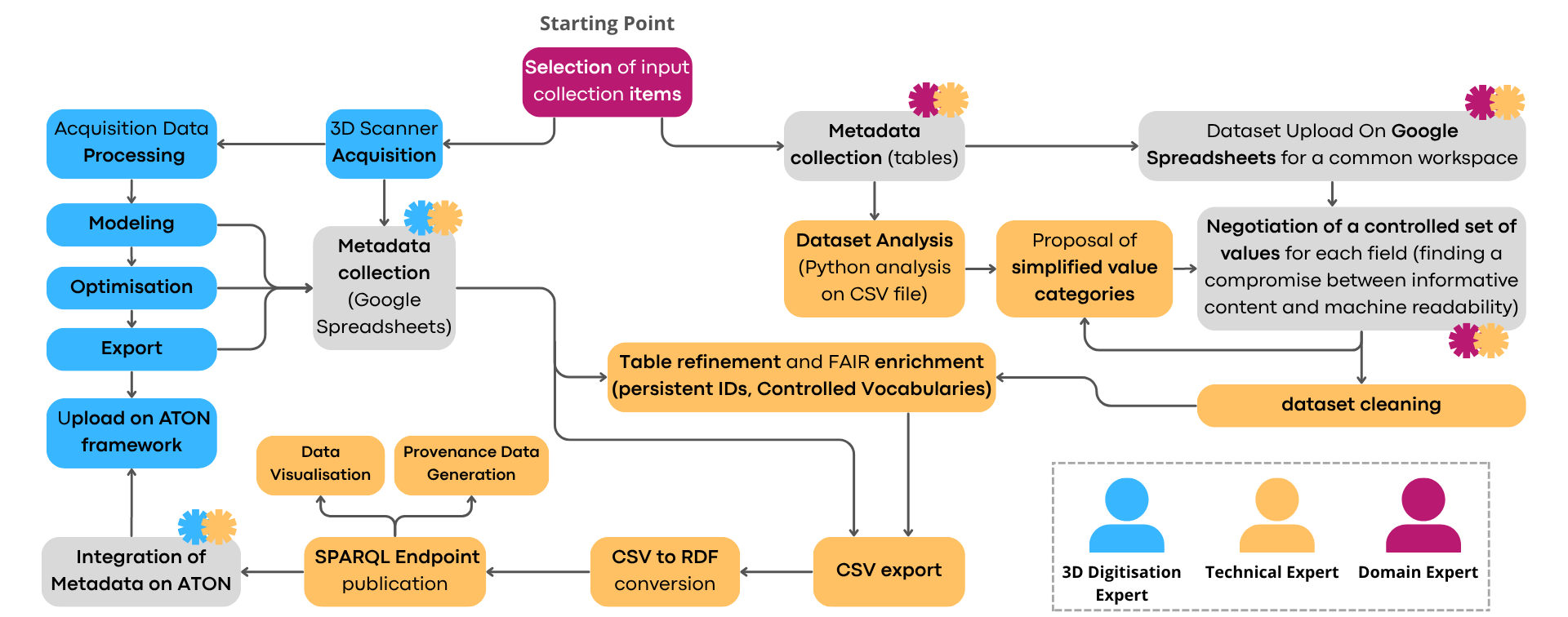}
\caption{The 3D digitisation process, an inherently multidisciplinary procedure, involving multiple actors with distinct roles and expertise (3D digitisation, technical, and domain). At each stage, one or several actors (grey boxes) are involved in the production of new data, making coordinated collaboration essential for effective metadata management.} \label{data-management}
\end{figure}

As illustrated in Figure~\ref{data-management}, two sets of metadata were collected with the help of the experts involved in the digitisation workflow: a dataset (bibliographic metadata, BM from now on) concerning the exhibited objects’ metadata, and another one on the digitisation process information (paradata, or PD from now on). Since the process of selecting, producing, collecting and managing metadata/paradata was meant to be collaborative, the chosen workspace needed to be accessible online by several actors at the same time. The group of involved experts was composed of people with different backgrounds and expertise, spanning from domain experts with deep knowledge concerning museum information, technicians and researchers involved in the digitisation process, and digital humanists acquainted with metadata management and Semantic Web Technologies. Thus, another requirement was having an intuitive platform, allowing for the customisation of the templates with features such as the limitation of data entries to controlled values, flexibility in the visualisations, and the definition of the accepted datatypes for specific fields. The choice fell on Google Sheets, as it optimally met all of the above-mentioned requirements. The templates on which the datasets were produced are available online in CSV and ODS formats and can be freely downloaded for replicating the collection process~\cite{moretti_2024_14277220}. Each of the two templates comes with an empty model and a tabular sample with a minimum set of values, providing at least one example for each field to fill. While currently the two templates are in Italian, they are accompanied by detailed documentation in Italian and English.

The process of semantic materialisation is based on both BM and PD, exported in CSV, to be transformed into a Turtle serialised RDF graph compliant with CHAD-AP. The complexity of the CSV structure and the intricacies of the values determined the necessity to address the conversion with a tool allowing for easy customisation. To keep the pipeline collaborative and accessible, an approach based on mapping languages was preferred over code-oriented alternatives.

In detail, based on these criteria, the RDF Mapping Language (RML)~\cite{dimou2014rml} was selected as the primary conversion technology. Indeed, RML is configurable through mapping rules defined in RDF files, does not require programming skills to be used for simple conversions, and provides a set of built-in functions to manipulate input datasets both in JSON and CSV format. Nonetheless, managing the Turtle configuration files still requires some acquaintance with Semantic Web technologies, and any potentially needed codebase expansion implies programming skills and the use of RDF serialisation files. To overcome these limitations, for the materialisation task was adopted Morph-KGC~\cite{arenas-guerrero_morph-kgc_2024}, a Python software tool based on RML. The advantages of adopting this approach include: easily extensible code by adding user-defined functions~\cite{arenas2024rml}, configuration files in the human-readable format YARRRML, a set of tools guiding and facilitating the definition of the mapping rules, a wide documentation, and an active GitHub community. Indeed, although a minimum acquaintance with the adopted data model is needed to express how to map the implicit relations of the input data structure to classes and properties, the tool remains accessible. Moreover, the online editor MATEY (\url{https://rml.io/yarrrml/matey/}) is made available to unexperienced users to get started with the conversion technology without the need to install local environments. 

\begin{figure}[h!]
\includegraphics[width=\linewidth]{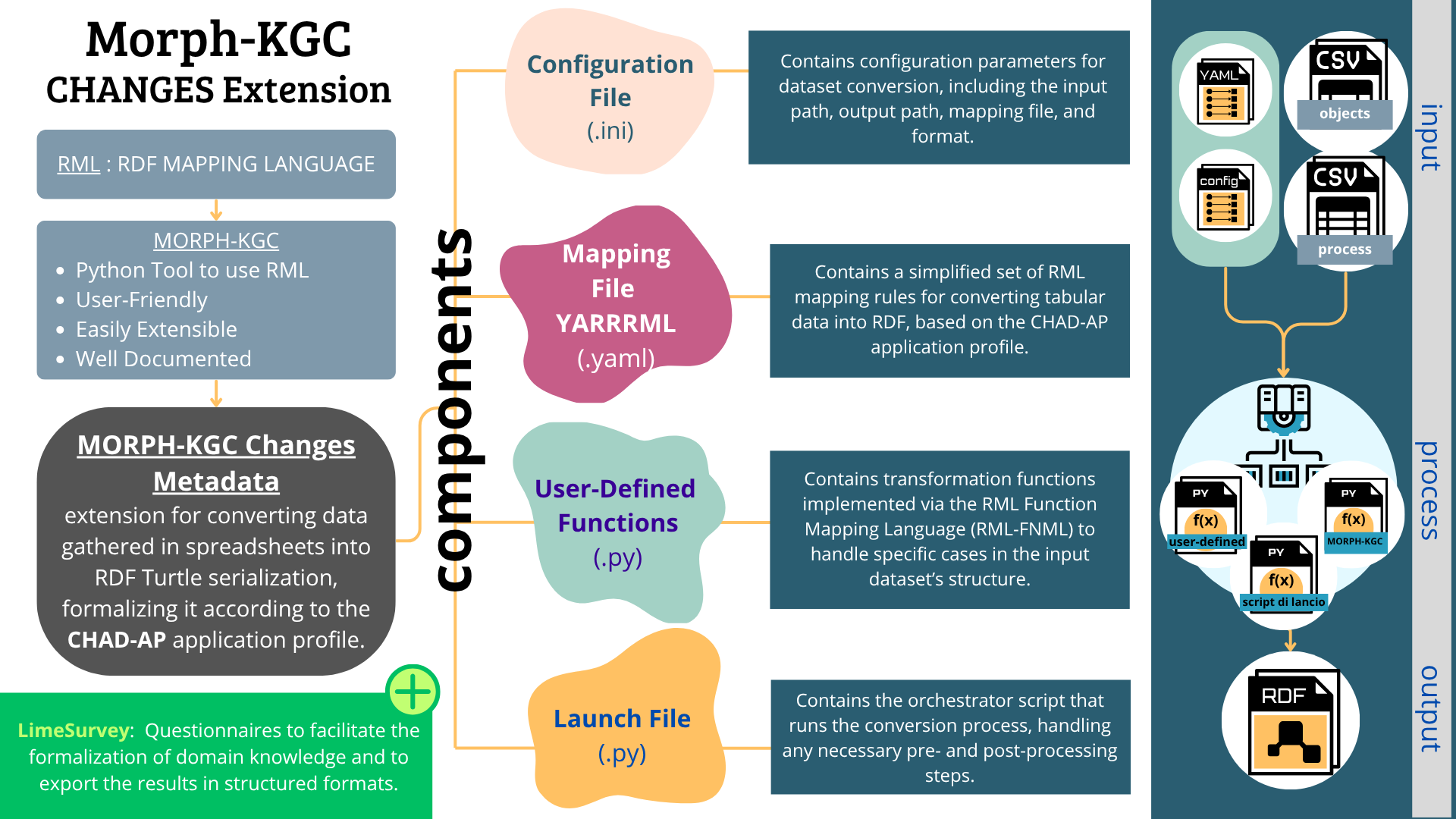}
\caption{The main components and additional customisations that are part of the Morph-KGC extension used to construct CHAD-KG.} \label{morph-kgc-components}
\end{figure}

Since the input datasets collected for the digitisation of the temporary exhibition presented specific complexities, their conversion was addressed in a Morph-KGC ad hoc software extension. The tool is open source and available on GitHub (\url{https://github.com/dharc-org/morph-kgc-changes-metadata}). The graph materialisation was thus based on the main components of the original software tool and customised additions (as illustrated in Figure~\ref{morph-kgc-components}). More in detail, the extended elements are listed below.

\subsubsection{Mapping Files.} Morph-KGC supports both Turtle and YARRRML mapping files. Consistent with the principles of the project, the most human-readable alternative was selected, and thus the mapping files were developed in YARRRML format. The rules for addressing the conversion of each input dataset template were managed in a separate file, defining how each CSV field should be translated into RDF, based on CHAD-AP classes and properties. As they are based on the structure of the template and not on the tables’ content, the mapping files can be reused to convert any other tabular dataset collected in the CHANGES tabular templates.

\subsubsection{Configuration file.} A configuration file was produced where the parameters for the tabular dataset conversion were declared. Among those, there are the input and output file paths, the desired RDF serialisation, and which mapping file to use. The configuration file allows for managing multiple conversions; in this case, for each of the datasets, a separate section should be declared with its specific parameters.

\subsubsection{User-defined functions.} Morph-KGC provides built-in declarative transformation functions implemented using the RML Function Mapping Language (RML-FNML)~\cite{arenas2024rml}, which translates the original RML functions declared in Java into Python. These functions are meant to address common value-extrapolation tasks, such as extracting multiple subjects from a unique data structure field, to propagate the relation expressed in the field to each of them. In case the intricacies of a dataset are not fully manageable with the set of built-in functions, it is possible to extend the codebase with additional user-defined functions. For the creation of the graph based on the two tabular input datasets, further functions were created, addressing specific tasks. In detail, the additional functions include: \texttt{normalize\_and\_convert\_to\_iri}, to create normalised IRIs of generated entities; \texttt{multiple\_separator\_split\_explode}, to handle a customisable range of separators and extract multiple values from complex fields; \texttt{assess\_aat\_tool\_type}, to map an acquisition technique to its Getty AAT code; \texttt{date\_to\_xs\_datetime}, to convert YYYY-MM-DD strings to \texttt{xsd:dateTime} literals; \texttt{split\_year\_range\_to\_dates}, to extract the start and end date from ranges of years; \texttt{convert\_to\_aat}, to map multiple object production techniques to Getty AAT codes; \texttt{convert\_documentary\_type\_to\_aat}, to map documentary types to AAT codes; \texttt{extract\_title}, to extract the string of a title by removing the language tag; \texttt{extract\_documented\_in\_iri}, to extract identifiers such as VIAF, ULAN or ORCID from complex strings containing the ID among other textual information; and \texttt{conditional\_normalize\_and\_convert\_to\_iri}, to generate IRIs only if the relationship type matches the intended one.

\subsubsection{Launch script.} The process is launched by executing a Python file that orchestrates the conversion process. Beyond this, this script also performs some pre- and post-processing tasks, including data normalisation, cleaning and needed reshaping to prepare the input for the RDF serialisation. In particular, it addresses the intricacies of the management of input datasets that might have structures significantly different from the expected one, in terms of missing or exceeding columns or values. Indeed, since some mapping rules involve multiple fields, the absence of one value can impede the materialisation of a full block of triples if not managed adequately. For this reason, the launch script was developed to be as flexible as possible, addressing a wide range of configurations, also including the absence of one of the two input datasets. In that case, a well-formed independent graph would be generated from the single input table, without causing blocking errors.

\subsection{Resource publishing and access}\label{publishing}

CHAD-KG is published following FAIR principles and made available through multiple access points. In addition to Zenodo, the data contained in CHAD-KG can be accessed through a public SPARQL endpoint (\url{https://w3id.org/dharc/sparql/chad-kg}), which enables users to perform queries directly on the graph using the SPARQL query language. In addition to that, the data is also made available through a static site (\url{https://w3id.org/dharc/web/chad-kg}) generated via the \emph{Static PUblisher of Knowledge} (SPUK)~\cite{barzaghi_sbrztspuk_2025}, a static site generator designed to easily publish a RDF KG as a simple HTML website. Also, SPUK provides the data in several RDF formats (N-triples, Turtle, JSON-LD, RDF/XML) to enable content negotiation of the entities in the KG via \emph{w3id.org}.

Thanks to SPUK, the KG can be explored through a user interface with a searchable list of entities. Each entity has its own HTML page displaying its properties and corresponding objects, whether they are other internal KG entities (linked to their respective pages), external references (linked to online sources), or literals (presented as plain text). SPUK also provides basic statistics about the KG, including the number of triples, entities, classes, and properties, with selected metrics visualised in charts. These highlight model usage, entity distribution across classes, and property frequency. Additionally, SPUK includes a browser-based SPARQL query interface (\url{https://w3id.org/dharc/sparql/chad-kg.html}), supporting both pre-defined queries and custom queries entered in a dedicated input area.

\section{Discussion}\label{discussion}

As mentioned in the sections before, in its current version openly available on Zenodo~\cite{moretti_2025_15102846} under a CC0 1.0 license, CHAD-KG comprises a total of 52,080 triples describing 14,506 entities (with an average of roughly 5 properties per entity), including the CHOs in the exhibition and all the accompanying entities that are part of the CHOs’ history, context and digitisation process (people, organisations, activities, tools, licenses, etc.).

\begin{figure}[h!]
\includegraphics[width=\linewidth]{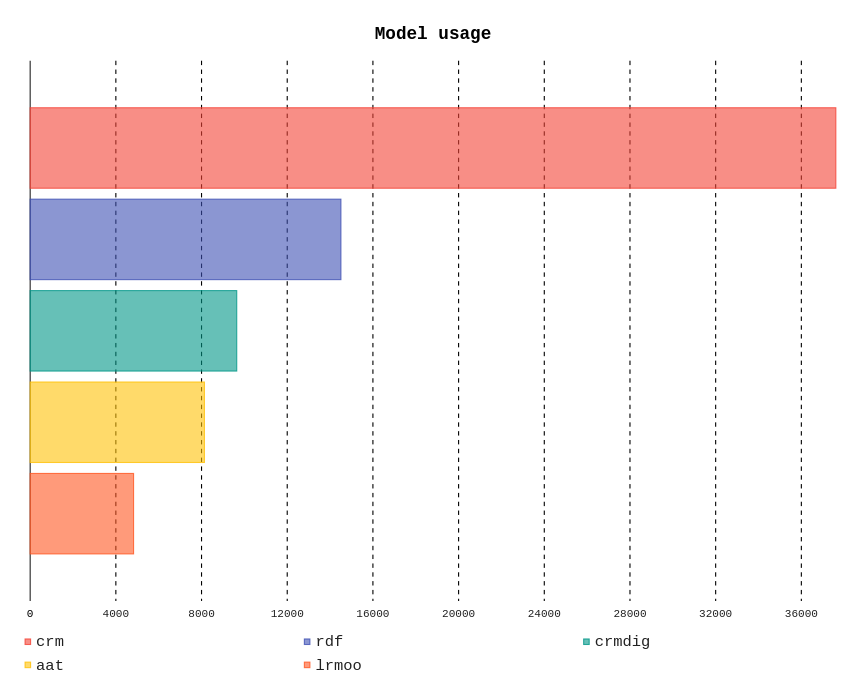}
\caption{The number of times each model is reused within CHAD-KG.} \label{model-usage}
\end{figure}

\begin{figure}[h!]
\includegraphics[width=\linewidth]{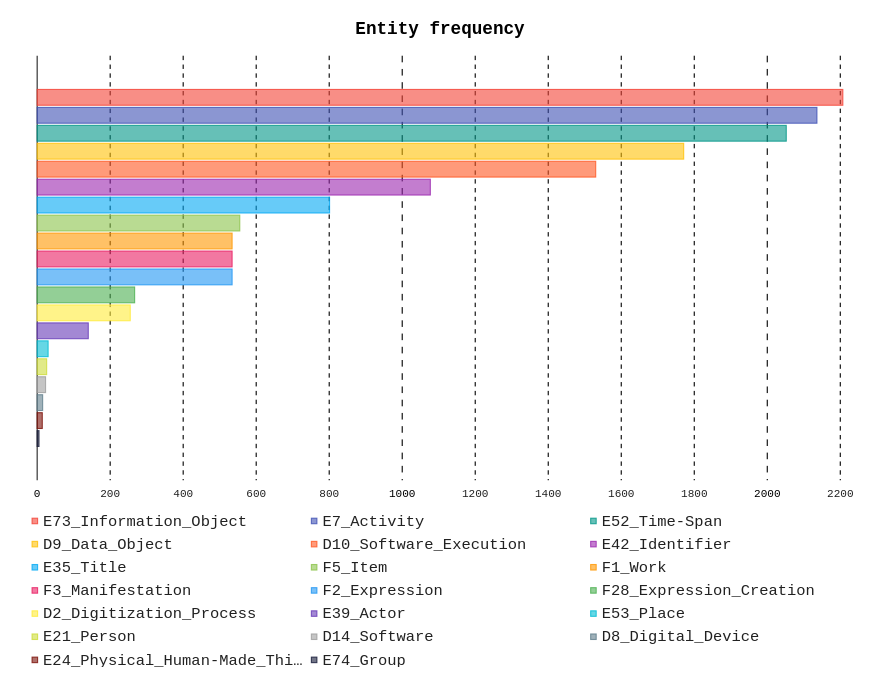}
\caption{The number of entities per class in CHAD-KG.} \label{entity-frequency}
\end{figure}

\begin{figure}[h!]
\includegraphics[width=\linewidth]{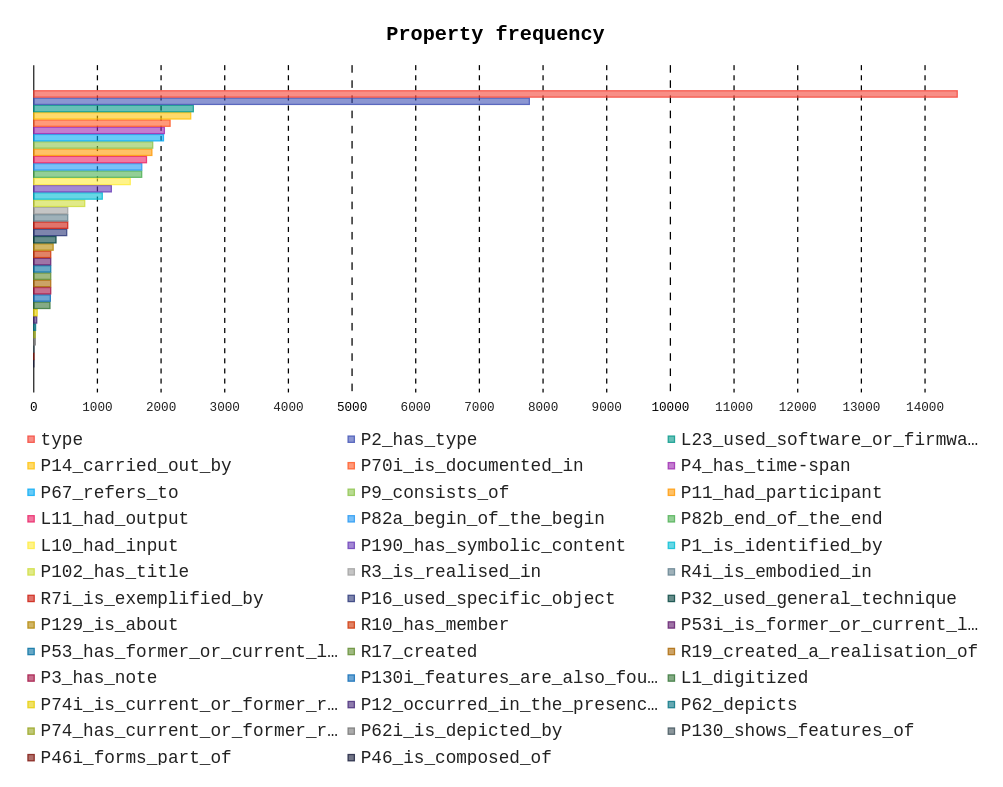}
\caption{The total count of properties occurrences in CHAD-KG.} \label{property-frequency}
\end{figure}

CHAD-KG is grounded in widely adopted  standards, including CIDOC-CRM, LRMoo, CRMdig, and AAT. It reuses 20 classes and 38 properties taken from CHAD-AP. Figure~\ref{model-usage} shows quantitatively how the models were reused in the KG, where the usage is based on the number of references to classes, entities and properties belonging to those models that occur within the graph. As expected, CIDOC-CRM is the most reused among all (37610 times for classes and properties), followed by CRMdig (9643), AAT (8135, for entities) and LRMoo (4827). This point is further illustrated in Figure~\ref{entity-frequency} and Figure~\ref{property-frequency}, which show respectively the frequencies of the entities per class and the total number of properties occurring in the graph.

In order to reach these results, the development of CHAD-KG has followed an iterative process, shaped by successive refinements of the underlying data model, the materialisation engine, and the KG itself. A key design goal of the project was to maintain a data production pipeline for KGs in the domain of cultural heritage digitisation that is sustainable, interoperable, reproducible and intelligible, in terms of conceptualisation, production and publishing. This brought us to construct CHAD-KG, which fills a critical gap in modelling metadata and paradata, offering a fine-grained account of how CHOs are transformed into digital surrogates. It already contributed to the advancement of Semantic Web practices in the cultural heritage domain, since it is being used as the single source of truth for the metadata displayed in the Digital Twin of the temporary exhibition \emph{The Other Renaissance - Ulisse Aldrovandi and The Wonders Of The World}, thus demonstrating real-world application and relevance to both the Semantic Web community and the wider heritage sector.

%Interestingly enough, CHAD-KG also showed us the value of a human-in-the-loop approach in producing a KG through a workflow that is based on collaboration between domain specialists, who are often not trained in technical aspects of semantic technologies, and technical experts, such as software developers and ontology engineers. By prioritising clarity and communicability over full automation or black-box solutions, we enabled a more transparent and participatory process.

\section{Conclusions}\label{conclusion}

In this paper, we described CHAD-KG, a new KG representing a temporary exhibition in terms of both the objects that were part of it and the processes that allowed for the acquisition and digitisation of those objects. Currently, it contains a rich set of RDF triples, produced via an extension of the Morph-KGC software and structured according to semantic patterns made possible through the use of CHAD-AP, a OWL-encoded application profile based on widely used models such as CIDOC-CRM, LRMoo, CRMdig and AAT. At the moment, CHAD-KG is being used as the single source of truth for the metadata and paradata displayed in an instance of the Aldrovandi Digital Twin that can be traversed and interacted with within a 3D virtual environment. Moreover, CHAD-KG is planned to include data from other eight case studies about digitisation initiatives of collections from various cultural heritage institutions that are involved in the CHANGES Project. In addition to that, the supporting technologies, such as CHAD-AP and the Morph-KGC extension, can (and are already planned to) be reused in similar projects centered around the digitisation of diverse CHOs branching outside our existing scope, from ceramics to historical posters collections. 

Despite the progress made so far, there is still much work to do. CHAD-AP shows some limitations in terms of modelling certain aspects of the cultural heritage domain, such as a stable mechanism to represent human-readable labels for certain entities (people, organisations, devices, concepts, etc.) that can be used reliably in frontend interfaces; the physical dimensions of a CHO (such as height, width and depth), expressed with specific values and measurement units; the materials a CHO is made of, such as canvas, stone, ceramic, etc.. While the Morph-KGC extension is accessible for case studies similar to those it was tested on, adding customisations for projects with specific needs requires programming skills and is not easily configurable otherwise.

In future work, we aim to further extend CHAD-AP capabilities in representing the domain of acquisition and digitisation for cultural heritage by applying it to other case studies. In addition to that, we aim to make the Morph-KGC extension more reusable via the implementation of a user-friendly interface that allows formalising RML rules needed for materialisation without expertise in coding nor Semantic Web technologies, such as a configurable LimeSurvey questionnaire, as suggested in~\cite{moretti_workflow_2024}.

\subsubsection{Resource availability statement.} A dump of CHAD-KG is freely available on Zenodo~\cite{moretti_2025_15102846} under a CC0 1.0 license. The SPARQL endpoint is available at \url{https://w3id.org/dharc/sparql/chad-kg}. The static site generated with SPUK is available at \url{https://w3id.org/dharc/web/chad-kg}. The current version of SPUK, released under a MIT license, is published on Zenodo~\cite{barzaghi_sbrztspuk_2025}. The most recent version of CHAD-AP, licensed under a CC BY 4.0 license, as well as its human-readable documentation, are publicly available on the w3id.org service at \url{https://w3id.org/dharc/ontology/chad-ap}. The repository containing the development documentation produced during SAMOD iterations is available on GitHub at \url{https://github.com/dharc-org/chad-ap/}, whereas an archived version of the repository can be found on Zenodo~\cite{barzaghi_dharc-orgchad-ap_2025}. The Morph-KGC extension is available on GitHub at \url{https://github.com/dharc-org/morph-kgc-changes-metadata} under an Apache v2.0 license.

\begin{credits}
\subsubsection{\ackname} This work has been funded by Project PE 0000020 CHANGES - CUP B53C22003780006, NRP Mission 4 Component 2 Investment 1.3, Funded by the European Union - NextGenerationEU.

\subsubsection{\discintname}
The authors have no competing interests to declare that are relevant to the content of this article.
\end{credits}

% ---- Bibliography ----
\bibliographystyle{splncs04}
\bibliography{bibliography}

\begin{thebibliography}{10}
\providecommand{\url}[1]{\texttt{#1}}
\providecommand{\urlprefix}{URL }
\providecommand{\doi}[1]{https://doi.org/#1}

\bibitem{angelis_chekg_2024}
Angelis, S., Moraitou, E., Caridakis, G., Kotis, K.: {CHEKG}: a collaborative and hybrid methodology for engineering modular and fair domain-specific knowledge graphs. Knowledge and Information Systems  \textbf{66}(8),  4899--4925 (Aug 2024). \doi{10.1007/s10115-024-02110-w}, \url{https://doi.org/10.1007/s10115-024-02110-w}

\bibitem{arenas2024rml}
Arenas-Guerrero, J., Espinoza-Arias, P., Bernab{\'e}-Diaz, J.A., Deshmukh, P., S{\'a}nchez-Fern{\'a}ndez, J.L., Corcho, O.: An {RML-FNML} module for {Python} user-defined functions in {Morph-KGC}. SoftwareX  \textbf{26},  101709 (2024)

\bibitem{arenas-guerrero_morph-kgc_2024}
Arenas-Guerrero, J., Chaves-Fraga, D., Toledo, J., Pérez, M.S., Corcho, O.: {Morph-KGC}: {Scalable} knowledge graph materialization with mapping partitions. Semantic Web  \textbf{15}(1),  1--20 (Jan 2024). \doi{10.3233/SW-223135}, \url{https://journals.sagepub.com/action/showAbstract}, publisher: SAGE Publications

\bibitem{balzani_saving_2024}
Balzani, R., Barzaghi, S., Bitelli, G., Bonifazi, F., Bordignon, A., Cipriani, L., Colitti, S., Collina, F., Daquino, M., Fabbri, F., Fanini, B., Fantini, F., Ferdani, D., Fiorini, G., Formia, E., Forte, A., Giacomini, F., Girelli, V.A., Gualandi, B., Heibi, I., Iannucci, A., Manganelli Del~Fà, R., Massari, A., Moretti, A., Peroni, S., Pescarin, S., Renda, G., Ronchi, D., Sullini, M., Tini, M.A., Tomasi, F., Travaglini, L., Vittuari, L.: Saving temporary exhibitions in virtual environments: {The} {Digital} {Renaissance} of {Ulisse} {Aldrovandi} – {Acquisition} and digitisation of cultural heritage objects. Digital Applications in Archaeology and Cultural Heritage  \textbf{32},  e00309 (Mar 2024). \doi{10.1016/j.daach.2023.e00309}, \url{https://www.sciencedirect.com/science/article/pii/S2212054823000541}

\bibitem{barbuti_creating_2020}
Barbuti, N.: Creating {Digital} {Cultural} {Heritage} with {Open} {Data}: {From} {FAIR} to {FAIR5} {Principles}. In: Ceci, M., Ferilli, S., Poggi, A. (eds.) Digital Libraries: The Era of Big Data and Data Science. pp. 173--181. Springer International Publishing, Cham (2020). \doi{10.1007/978-3-030-39905-4_17}

\bibitem{barzaghi_sbrztspuk_2025}
Barzaghi, S.: {SPUK}: v0.1.0 (May 2025). \doi{10.5281/zenodo.15389756}, \url{https://zenodo.org/records/15389756}

\bibitem{barzaghi_proposal_2024}
Barzaghi, S., Bordignon, A., Gualandi, B., Heibi, I., Massari, A., Moretti, A., Peroni, S., Renda, G.: A {Proposal} for a {FAIR} {Management} of {3D} {Data} in {Cultural} {Heritage}: {The} {Aldrovandi} {Digital} {Twin} {Case}. Data Intelligence  \textbf{6}(4),  1190--1221 (Dec 2024). \doi{10.3724/2096-7004.di.2024.0061}, \url{https://www.sciengine.com/doi/10.3724/2096-7004.di.2024.0061}

\bibitem{barzaghi_developing_2025}
Barzaghi, S., Heibi, I., Moretti, A., Peroni, S.: Developing {Application} {Profiles} for {Enhancing} {Data} and {Workflows} in {Cultural} {Heritage} {Digitisation} {Processes}. In: Demartini, G., Hose, K., Acosta, M., Palmonari, M., Cheng, G., Skaf-Molli, H., Ferranti, N., Hernández, D., Hogan, A. (eds.) The Semantic Web – ISWC 2024. pp. 197--217. Springer Nature Switzerland, Cham (2025). \doi{10.1007/978-3-031-77847-6_11}

\bibitem{barzaghi_dharc-orgchad-ap_2025}
Barzaghi, S., Moretti, A., Peroni, S.: {CHAD-AP}: v2.0.1 (May 2025). \doi{10.5281/zenodo.15391304}, \url{https://doi.org/10.5281/zenodo.15391304}

\bibitem{battle_enabling_2012}
Battle, R., Kolas, D.: Enabling the geospatial {Semantic} {Web} with {Parliament} and {GeoSPARQL}. Semantic Web  \textbf{3}(4),  355--370 (Nov 2012). \doi{10.3233/SW-2012-0065}, \url{https://journals.sagepub.com/action/showAbstract}, publisher: SAGE Publications

\bibitem{carriero_arco_2019}
Carriero, V.A., Gangemi, A., Mancinelli, M.L., Marinucci, L., Nuzzolese, A.G., Presutti, V., Veninata, C.: {ArCo}: {The} {Italian} {Cultural} {Heritage} {Knowledge} {Graph}. In: Ghidini, C., Hartig, O., Maleshkova, M., Svátek, V., Cruz, I., Hogan, A., Song, J., Lefrançois, M., Gandon, F. (eds.) The Semantic Web – ISWC 2019. pp. 36--52. Springer International Publishing, Cham (2019). \doi{10.1007/978-3-030-30796-7_3}

\bibitem{desul_semantic_2023}
Desul, S., Mahapatra, R.K., Patra, R.K., Sethy, M., Pandey, N.: Semantic technology for cultural heritage: a bibliometric-based review. Global Knowledge, Memory and Communication  \textbf{ahead-of-print}(ahead-of-print) (Aug 2023). \doi{10.1108/GKMC-04-2023-0125}, \url{https://www.emerald.com/insight/content/doi/10.1108/gkmc-04-2023-0125/full/html}, publisher: Emerald Publishing Limited

\bibitem{dimou2014rml}
Dimou, A., Vander~Sande, M., Colpaert, P., Verborgh, R., Mannens, E., Van~de Walle, R.: {RML}: {A} generic language for integrated {RDF} mappings of heterogeneous data. Ldow  \textbf{1184} (2014)

\bibitem{doerr_cidoc_2003}
Doerr, M.: The {CIDOC} {Conceptual} {Reference} {Module}: {An} {Ontological} {Approach} to {Semantic} {Interoperability} of {Metadata}. AI Magazine  \textbf{24}(3), ~75 (Sep 2003). \doi{10.1609/aimag.v24i3.1720}, \url{https://ojs.aaai.org/aimagazine/index.php/aimagazine/article/view/1720}, section: Articles

\bibitem{doerr_ontologies_2009}
Doerr, M.: Ontologies for {Cultural} {Heritage}. In: Staab, S., Studer, R. (eds.) Handbook on {Ontologies}, pp. 463--486. Springer, Berlin, Heidelberg (2009). \doi{10.1007/978-3-540-92673-3_21}, \url{https://doi.org/10.1007/978-3-540-92673-3_21}

\bibitem{doerr2014crmsci}
Doerr, M., Kritsotaki, A., Rousakis, Y., Hiebel, G., Theodoridou, M.: {CRMsci}: {The} scientific observation model. Tech. rep., FORTH, Tech. Rep., 2014.[Online]. Available: https://projects.ics.forth.gr/isl/CRMext/CRMsci/docs/CRMsci1.1.pdf (2014), \url{https://projects.ics.forth.gr/isl/CRMext/CRMsci/docs/CRMsci1.1.pdf}

\bibitem{doerr2011crmdig}
Doerr, M., Theodoridou, M.: {CRMdig}: {A} generic digital provenance model for scientific observation. In: 3rd USENIX Workshop on the Theory and Practice of Provenance (TaPP 11) (2011), \url{https://www.usenix.org/legacy/events/tapp11/tech/final_files/Doerr.pdf}

\bibitem{gangemi_ontology_2009}
Gangemi, A., Presutti, V.: Ontology {Design} {Patterns}. In: Staab, S., Studer, R. (eds.) Handbook on Ontologies, pp. 221--243. Springer, Berlin, Heidelberg (2009). \doi{10.1007/978-3-540-92673-3_10}, \url{https://doi.org/10.1007/978-3-540-92673-3_10}

\bibitem{grant_reusable_2023}
Grant, R.: Reusable, {FAIR} {Humanities} {Data}: {Creating} {Practical} {Guidance} for {Authors} at {Routledge} {Open} {Research}. International Journal of Digital Curation  \textbf{17}(1),  15--15 (2023). \doi{10.2218/ijdc.v17i1.820}, \url{https://www.ijdc.net/index.php/ijdc/article/view/820}, number: 1

\bibitem{gualandi_what_2022}
Gualandi, B., Pareschi, L., Peroni, S.: What do we mean by “data”? {A} proposed classification of data types in the arts and humanities. Journal of Documentation  \textbf{79}(7),  51--71 (Dec 2022). \doi{10.1108/JD-07-2022-0146}, \url{https://www.emerald.com/insight/content/doi/10.1108/jd-07-2022-0146/full/html}, publisher: Emerald Publishing Limited

\bibitem{guizzardi_ontology_2020}
Guizzardi, G.: Ontology, {Ontologies} and the “{I}” of {FAIR}. Data Intelligence  \textbf{2}(1-2),  181--191 (Jan 2020). \doi{10.1162/dint_a_00040}, \url{https://doi.org/10.1162/dint_a_00040}

\bibitem{han_opennre_2019}
Han, X., Gao, T., Yao, Y., Ye, D., Liu, Z., Sun, M.: {OpenNRE}: {An} {Open} and {Extensible} {Toolkit} for {Neural} {Relation} {Extraction} (Sep 2019). \doi{10.48550/arXiv.1909.13078}, \url{http://arxiv.org/abs/1909.13078}, arXiv:1909.13078 [cs]

\bibitem{hao_construction_2021}
Hao, X., Ji, Z., Li, X., Yin, L., Liu, L., Sun, M., Liu, Q., Yang, R.: Construction and {Application} of a {Knowledge} {Graph}. Remote Sensing  \textbf{13}(13), ~2511 (Jan 2021). \doi{10.3390/rs13132511}, \url{https://www.mdpi.com/2072-4292/13/13/2511}, number: 13 Publisher: Multidisciplinary Digital Publishing Institute

\bibitem{harpring_development_2010}
Harpring, P.: Development of the {Getty} {Vocabularies}: {AAT}, {TGN}, {ULAN}, and {CONA}. Art Documentation: Journal of the Art Libraries Society of North America  \textbf{29}(1),  67--72 (Apr 2010). \doi{10.1086/adx.29.1.27949541}, \url{https://www.journals.uchicago.edu/doi/abs/10.1086/adx.29.1.27949541}, publisher: The University of Chicago Press

\bibitem{harrower_sustainable_2020}
Harrower, N., Immenhauser, B., Lauer, G., Maryl, M., Orlandi, T., Rentier, B., Wandl-Vogt, E.: Sustainable and {FAIR} {Data} {Sharing} in the {Humanities} (Feb 2020). \doi{10.7486/DRI.tq582c863}, \url{https://orbi.uliege.be/handle/2268/246860}, publisher: ALLEA - All European Academies, Berlin, Germany

\bibitem{haslhofer_dataeuropeanaeu_2011}
Haslhofer, B., Isaac, A.: data.europeana.eu: {The} {Europeana} {Linked} {Open} {Data} {Pilot}. In: Proceedings of the International Conference on Dublin Core and Metadata Applications. Dublin Core Metadata Initiative (Sep 2011). \doi{10.23106/dcmi.952135673}, \url{https://dcpapers.dublincore.org/article/952135673}

\bibitem{hiebel_ontological_2021}
Hiebel, G., Aspöck, E., Kopetzky, K.: Ontological {Modeling} for {Excavation} {Documentation} and {Virtual} {Reconstruction} of an {Ancient} {Egyptian} {Site}. J. Comput. Cult. Herit.  \textbf{14}(3),  32:1--32:14 (Jul 2021). \doi{10.1145/3439735}, \url{https://dl.acm.org/doi/10.1145/3439735}

\bibitem{homburg_metadata_2021}
Homburg, T., Cramer, A., Raddatz, L., Mara, H.: Metadata schema and ontology for capturing and processing of {3D} cultural heritage objects. Heritage Science  \textbf{9}(1),  1--19 (Jul 2021). \doi{10.1186/s40494-021-00561-w}, \url{https://www.nature.com/articles/s40494-021-00561-w}, publisher: Nature Publishing Group

\bibitem{hyvonen_knowledge-based_2019}
Hyvönen, E., Rantala, H.: Knowledge-based {Relation} {Discovery} in {Cultural} {Heritage} {Knowledge} {Graphs}. Digital Humanities in the Nordic and Baltic Countries Publications  \textbf{2}(1),  230--239 (May 2019). \doi{10.5617/dhnbpub.11098}, \url{https://journals.uio.no/dhnbpub/article/view/11098}

\bibitem{ioannides_integrating_2025}
Ioannides, M., Karittevli, E., Panayiotou, P., Baker, D.: Integrating {Paradata}, {Metadata}, and {Data} for an {Effective} {Memory} {Twin} in the {Field} of {Digital} {Cultural} {Heritage}. In: Ioannides, M., Baker, D., Agapiou, A., Siegkas, P. (eds.) 3D Research Challenges in Cultural Heritage V: Paradata, Metadata and Data in Digitisation, pp. 24--35. Springer Nature Switzerland, Cham (2025). \doi{10.1007/978-3-031-78590-0_3}, \url{https://doi.org/10.1007/978-3-031-78590-0_3}

\bibitem{jain_domain-specific_2020}
Jain, N.: {Domain-Specific} {Knowledge} {Graph} {Construction} for {Semantic} {Analysis}. In: Harth, A., Presutti, V., Troncy, R., Acosta, M., Polleres, A., Fernández, J.D., Xavier~Parreira, J., Hartig, O., Hose, K., Cochez, M. (eds.) The Semantic Web: ESWC 2020 Satellite Events. pp. 250--260. Springer International Publishing, Cham (2020). \doi{10.1007/978-3-030-62327-2_40}

\bibitem{khorashadizadeh_research_2024}
Khorashadizadeh, H., Amara, F.Z., Ezzabady, M., Ieng, F., Tiwari, S., Mihindukulasooriya, N., Groppe, J., Sahri, S., Benamara, F., Groppe, S.: Research {Trends} for the {Interplay} between {Large} {Language} {Models} and {Knowledge} {Graphs} (Aug 2024). \doi{10.48550/arXiv.2406.08223}, \url{http://arxiv.org/abs/2406.08223}, arXiv:2406.08223 [cs]

\bibitem{larsson_developing_2025}
Larsson, a.M., Bornsäter, B., Hacke, M.: Developing practices for {FAIR} and linked data in {Heritage} {Science}. NPJ HERITAGE SCIENCE  \textbf{13}(1) (2025), \url{https://urn.kb.se/resolve?urn=urn:nbn:se:uu:diva-552424}, publisher: Springer Nature

\bibitem{lebo_prov-o_2013}
Lebo, T., Sahoo, S., McGuinness, D., Belhajjame, K., Cheney, J., Corsar, D., Garijo, D., Soiland-Reyes, S., Zednik, S., Zhao, J.: {PROV-O}: {The} {PROV} {Ontology} (Apr 2013), \url{https://research.manchester.ac.uk/en/publications/prov-o-the-prov-ontology}, publisher: World Wide Web Consortium

\bibitem{lehmann_dbpedia_2015}
Lehmann, J., Isele, R., Jakob, M., Jentzsch, A., Kontokostas, D., Mendes, P.N., Hellmann, S., Morsey, M., van Kleef, P., Auer, S., Bizer, C.: {DBpedia} – {A} large-scale, multilingual knowledge base extracted from {Wikipedia}. Semantic Web  \textbf{6}(2),  167--195 (Mar 2015). \doi{10.3233/SW-140134}, \url{https://journals.sagepub.com/action/showAbstract}, publisher: SAGE Publications

\bibitem{mero241_clariah_2020}
Mero\&\#241, o~Pe\&\#241, Uela, A., De~Boer, V., Van~Erp, M., Zijdeman, R., Mourits, R., Melder, W., Rijpma, A., Schalk, R.: {CLARIAH}: {Enabling} {Interoperability} {Between} {Humanities} {Disciplines} with {Ontologies}. In: Applications and Practices in Ontology Design, Extraction, and Reasoning, pp. 73--90. IOS Press (2020). \doi{10.3233/SSW200036}, \url{https://ebooks.iospress.nl/doi/10.3233/SSW200036}

\bibitem{moretti_2024_14277220}
Moretti, A., Barzaghi, S.: Modelli spreadsheet changes - acquisizione e oggetti (Dec 2024). \doi{10.5281/zenodo.14277220}, \url{https://doi.org/10.5281/zenodo.14277220}

\bibitem{moretti_2025_15102846}
Moretti, A., Barzaghi, S.: {CHAD-KG}: {TTL} {Serialised} {RDF} {Dataset} of {Exhibited} objects and digitisation process (Mar 2025). \doi{10.5281/zenodo.15102846}, \url{https://doi.org/10.5281/zenodo.15102846}

\bibitem{moretti_workflow_2024}
Moretti, A., Heibi, I., Peroni, S.: A {Workflow} for {GLAM} {Metadata} {Crosswalk} (May 2024). \doi{10.48550/arXiv.2405.02113}, \url{http://arxiv.org/abs/2405.02113}, arXiv:2405.02113 [cs]

\bibitem{padfield_semantic_2019}
Padfield, J., Kontiza, K., Bikakis, A., Vlachidis, A.: {Semantic} {Representation} and {Location} {Provenance} of {Cultural} {Heritage} {Information}: the {National} {Gallery} {Collection} in {London}. Heritage  \textbf{2}(1),  648--665 (Mar 2019). \doi{10.3390/heritage2010042}, \url{https://www.mdpi.com/2571-9408/2/1/42}, number: 1 Publisher: Multidisciplinary Digital Publishing Institute

\bibitem{pellegrino_move_2022}
Pellegrino, M.A., Scarano, V., Spagnuolo, C.: Move cultural heritage knowledge graphs in everyone’s pocket. Semantic Web  \textbf{14}(2),  323--359 (Dec 2022). \doi{10.3233/SW-223117}, \url{https://journals.sagepub.com/action/showAbstract}, publisher: SAGE Publications

\bibitem{perez-arriaga_construction_2018}
Perez-Arriaga, M.O., Estrada, T., Abad-Mota, S.: Construction of {Semantic} {Data} {Models}. In: Filipe, J., Bernardino, J., Quix, C. (eds.) Data Management Technologies and Applications. pp. 46--66. Springer International Publishing, Cham (2018). \doi{10.1007/978-3-319-94809-6_3}

\bibitem{peroni_simplified_2017}
Peroni, S.: A {Simplified} {Agile} {Methodology} for {Ontology} {Development}. In: Dragoni, M., Poveda-Villalón, M., Jimenez-Ruiz, E. (eds.) OWL: Experiences and Directions – Reasoner Evaluation. pp. 55--69. Springer International Publishing, Cham (2017). \doi{10.1007/978-3-319-54627-8_5}

\bibitem{peroni_food_2016}
Peroni, S., Lodi, G., Asprino, L., Gangemi, A., Presutti, V.: {FOOD}: {FOod} in {Open} {Data}. In: Groth, P., Simperl, E., Gray, A., Sabou, M., Krötzsch, M., Lecue, F., Flöck, F., Gil, Y. (eds.) The Semantic Web – ISWC 2016. pp. 168--176. Springer International Publishing, Cham (2016). \doi{10.1007/978-3-319-46547-0_18}

\bibitem{poljak_bilic_fairness_2024}
Poljak~Bilić, L., Posavec, K.: {FAIRness} of {Research} {Data} in the {European} {Humanities} {Landscape}. Publications  \textbf{12}(1), ~6 (Mar 2024). \doi{10.3390/publications12010006}, \url{https://www.mdpi.com/2304-6775/12/1/6}, number: 1 Publisher: Multidisciplinary Digital Publishing Institute

\bibitem{rahman_knowledge_2024}
Rahman, S., Choi, F., Kim, H., Zhang, D., Hruschka, E.: Knowledge {Acquisition} and {Integration} with {Expert-in-the-loop} (Feb 2024). \doi{10.48550/arXiv.2402.03291}, \url{http://arxiv.org/abs/2402.03291}, arXiv:2402.03291 [cs]

\bibitem{rajbhandari_agrovoc_2012}
Rajbhandari, S., Keizer, J.: The {AGROVOC} {Concept} {Scheme} – {A} {Walkthrough}. Journal of Integrative Agriculture  \textbf{11}(5),  694--699 (May 2012). \doi{10.1016/S2095-3119(12)60058-6}, \url{https://www.sciencedirect.com/science/article/pii/S2095311912600586}

\bibitem{ristoski_large-scale_2020}
Ristoski, P., Gentile, A.L., Alba, A., Gruhl, D., Welch, S.: Large-scale relation extraction from web documents and knowledge graphs with human-in-the-loop. Journal of Web Semantics  \textbf{60},  100546 (Jan 2020). \doi{10.1016/j.websem.2019.100546}, \url{https://www.sciencedirect.com/science/article/pii/S1570826819300782}

\bibitem{riva_lrmoo_2022}
Riva, P., Žumer, M., Aalberg, T.: {LRMoo}, a high-level model in an object-oriented framework (Oct 2022), \url{https://2022.ifla.org/}, publisher: International Federation of Library Associations and Institutions (IFLA)

\bibitem{schleider_silknow_2021}
Schleider, T., Troncy, R., Gaitán, M., Alba, E., Sebastián, J., Mladenic, D., Kastelic, A., Massri, M.B., León, A., Puren, M., Vernus, P., Clermont, D., Rottensteiner, F., Vitella, M., Cicero, G.L.: The {SILKNOW} {Knowledge} {Graph}. Semantic Web  (2021)

\bibitem{stocker_skg4eosc_2022}
Stocker, M., Heger, T., Schweidtmann, A., Ćwiek Kupczyńska, H., Penev, L., Dojchinovski, M., Willighagen, E., Vidal, M.E., Turki, H., Balliet, D., Tiddi, I., Kuhn, T., Mietchen, D., Karras, O., Vogt, L., Hellmann, S., Jeschke, J., Krajewski, P., Auer, S.: {SKG4EOSC} - {Scholarly} {Knowledge} {Graphs} for {EOSC}: {Establishing} a backbone of knowledge graphs for {FAIR} {Scholarly} {Information} in {EOSC}. Research Ideas and Outcomes  \textbf{8},  e83789 (Mar 2022). \doi{10.3897/rio.8.e83789}, \url{https://riojournal.com/article/83789/}, publisher: Pensoft Publishers

\bibitem{thanos_research_2017}
Thanos, C.: Research {Data} {Reusability}: {Conceptual} {Foundations}, {Barriers} and {Enabling} {Technologies}. Publications  \textbf{5}(1), ~2 (Mar 2017). \doi{10.3390/publications5010002}, \url{https://www.mdpi.com/2304-6775/5/1/2}, number: 1 Publisher: Multidisciplinary Digital Publishing Institute

\bibitem{toth-czifra_risk_2019}
Tóth-Czifra, E.: The risk of losing thick description: {Data} management challenges {Arts} and {Humanities} face in the evolving {FAIR} data ecosystem (Apr 2019), \url{https://shs.hal.science/halshs-02115505}

\bibitem{van_assche_balancing_2022}
Van~Assche, D.: Balancing {RDF} {Generation} from {Heterogeneous} {Data} {Sources}. In: Groth, P., Rula, A., Schneider, J., Tiddi, I., Simperl, E., Alexopoulos, P., Hoekstra, R., Alam, M., Dimou, A., Tamper, M. (eds.) The Semantic Web: ESWC 2022 Satellite Events. pp. 264--274. Springer International Publishing, Cham (2022). \doi{10.1007/978-3-031-11609-4_40}

\bibitem{vrandecic_wikidata_2014}
Vrandečić, D., Krötzsch, M.: Wikidata: a free collaborative knowledgebase. Communications of the ACM  \textbf{57}(10),  78--85 (Sep 2014). \doi{10.1145/2629489}, \url{https://dl.acm.org/doi/10.1145/2629489}

\bibitem{wilkinson_fair_2016}
Wilkinson, M.D., Dumontier, M., Aalbersberg, I.J., Appleton, G., Axton, M., Baak, A., Blomberg, N., Boiten, J.W., da~Silva~Santos, L.B., Bourne, P.E., Bouwman, J., Brookes, A.J., Clark, T., Crosas, M., Dillo, I., Dumon, O., Edmunds, S., Evelo, C.T., Finkers, R., Gonzalez-Beltran, A., Gray, A.J.G., Groth, P., Goble, C., Grethe, J.S., Heringa, J., ’t Hoen, P.A.C., Hooft, R., Kuhn, T., Kok, R., Kok, J., Lusher, S.J., Martone, M.E., Mons, A., Packer, A.L., Persson, B., Rocca-Serra, P., Roos, M., van Schaik, R., Sansone, S.A., Schultes, E., Sengstag, T., Slater, T., Strawn, G., Swertz, M.A., Thompson, M., van~der Lei, J., van Mulligen, E., Velterop, J., Waagmeester, A., Wittenburg, P., Wolstencroft, K., Zhao, J., Mons, B.: The {FAIR} {Guiding} {Principles} for scientific data management and stewardship. Scientific Data  \textbf{3}(1),  160018 (Mar 2016). \doi{10.1038/sdata.2016.18}, \url{https://www.nature.com/articles/sdata201618}, publisher: Nature Publishing Group

\bibitem{yang_knowledge_2023}
Yang, S., Hou, M.: Knowledge graph representation method for semantic {3D} modeling of {Chinese} grottoes. Heritage Science  \textbf{11}(1),  1--26 (Dec 2023). \doi{10.1186/s40494-023-01084-2}, \url{https://www.nature.com/articles/s40494-023-01084-2}, publisher: Nature Publishing Group

\bibitem{zhong_comprehensive_2023}
Zhong, L., Wu, J., Li, Q., Peng, H., Wu, X.: A {Comprehensive} {Survey} on {Automatic} {Knowledge} {Graph} {Construction}. ACM Comput. Surv.  \textbf{56}(4),  94:1--94:62 (Nov 2023). \doi{10.1145/3618295}, \url{https://dl.acm.org/doi/10.1145/3618295}

\bibitem{zumer_ifla_2018}
Žumer, M.: {IFLA} {Library} {Reference} {Model} ({IFLA} {LRM}) — {Harmonisation} of the {FRBR} {Family}. KNOWLEDGE ORGANIZATION  \textbf{45}(4),  310--318 (2018). \doi{10.5771/0943-7444-2018-4-310}, \url{https://www.imrpress.com/journal/KO/45/4/10.5771/0943-7444-2018-4-310}

\end{thebibliography}

\end{document}